\documentclass[a4paper,11pt]{article}

\usepackage{jinstpub} 

\title{Performance of the wavelength-shifting fiber upgrade for the Mu2e cosmic-ray veto detector}

\author[a]{D. Coveyou,}
\author[a]{E.C. Dukes,}
\author[a]{R.C. Group,}
\author[b]{Y. Oksuzian,}
\author[a]{S. Roberts,}
\author[a]{M. Solt$^1,$\note{Corresponding author.}}

\affiliation[a]{University of Virginia, Charlottesville, VA 22904, USA}
\affiliation[b]{Argonne National Laboratory, Argonne, IL 60439, USA}

\emailAdd{gtf9nz@virginia.edu}

\abstract{The Cosmic-Ray Veto detector for the muon-to-electron conversion experiment at Fermilab consists of four plastic scintillating counter layers read out by silicon photo-multipliers through embedded wavelength-shifting fibers. In order to increase the light yield in the most critical regions of the Cosmic-Ray Veto detector for improved background rejection, a 1.8\,mm diameter fiber is being used in many of the detector's critical modules instead of the previously planned 1.4\,mm diameter fiber. This paper reports the testing procedure and light properties of thirty-four 1.8\,mm fiber spools, with measurements performed using a custom-built scanner. We compare these new results with previously published data from the 1.4\,mm diameter fiber used for regions of the cosmic-ray veto where the increased light yield is not required. In addition, measurements of fiber aging were performed.}


\begin{document}
\maketitle
\flushbottom

\section{Introduction}
\label{sec:intro}

The muon-to-electron conversion (Mu2e) experiment~\cite{tdr,Abusalma:2018xem,Mu2e:2022ggl} will search for charged lepton flavor violation (CLFV) through the neutrinoless conversion of a muon into an electron in the presence of a nucleus. Muon-to-electron conversion occurs because of neutrino oscillations, but at an unobservable rate (approximately forty orders of magnitude below the current experimental limits). However, many extensions to the Standard Model of particle physics predict that this process could occur at an observable rate~\cite{Bernstein:2013hba}. An observation of this conversion process would be an unambiguous sign of physics beyond the Standard Model. 

To search for CLFV with the desired sensitivity, Mu2e must keep the total background to fewer than one event over the full multi-year experimental run. However, without a cosmic-ray veto system, approximately one background event will be produced per day at Mu2e from cosmic-ray muons. Thus, for the desired background suppression, Mu2e needs to reduce the cosmic-ray background by four orders of magnitude. By surrounding much of the Mu2e apparatus, the Cosmic-Ray Veto (CRV) is designed to veto cosmic-ray backgrounds with an overall 99.99\% efficiency. 

The CRV consists of plastic scintillator counters up to 7\,m long containing embedded wavelength-shifting (WLS) fibers that are read out by silicon photo-multipliers (SiPMs). The CRV efficiency critically depends on the performance of the fiber. Therefore, a WLS fiber scanner was designed and fabricated to measure fiber light yield properties and ensure the CRV's fiber quality requirements. More information about the fiber scanner and 1.4\,mm production fiber performance is given in previous publications~\cite{Dukes:2018scs, proc}. 

In order to improve cosmic ray background rejection efficiency, larger 1.8 mm diameter fiber will be used for the most critical modules of the CRV. Using the Fermilab test beam facility, the light-yield improvement of counters outfitted with 1.8\,mm diameter fiber was measured to be 24\%~\cite{TestBeam2018}. This light yield increase is consistent with a simple geometric argument that a fiber should absorb an amount of light proportional to its diameter ($\frac{1.8 \ \mathrm{mm}}{1.4 \ \mathrm{mm}} \sim 29\%$). To ensure sufficient light yield for the CRV's sectors with the largest flux of cosmic ray muons, 14\,km of larger diameter fiber was purchased. This paper reviews important fiber characteristics, describes the fiber scanner, and presents studies measuring the 1.8\,mm diameter fiber quality, including preliminary fiber aging studies.

\section{Wavelength-shifting fiber}
The CRV uses Kuraray Y11 non-S-type WLS fibers~\cite{kuraray} containing fluorescent dye, K27, that absorbs blue light (375-475\,nm) from scintillating counters and emits light in the green (450-600\,nm) spectral region. These fiber signals are read out by $2.0{\times}2.0$\,mm$^2$ (model S13360-2050VE with a pixel size of 50\,$\mu$m) Hamamatsu silicon photo-multipliers (SiPMs)~\cite{SiPM_rad}. The chosen SiPM spectral photon detection efficiency response is well-matched to the emission spectrum for Kuraray Y11 fiber.\footnote{
SiPMs in the CRV feature a peak sensitivity at 450\,nm, with 80\% (50\%) peak sensitivity at 550\,nm (620\,nm) \cite{sipm}.} Multi-clad non-S-type fiber was selected for the CRV due to its enhanced light yield and longer attenuation length~\cite{kuraray}. 

\section{Fiber scanner}
\begin{figure}[htb]
\centering 
\includegraphics[height=2.3in]{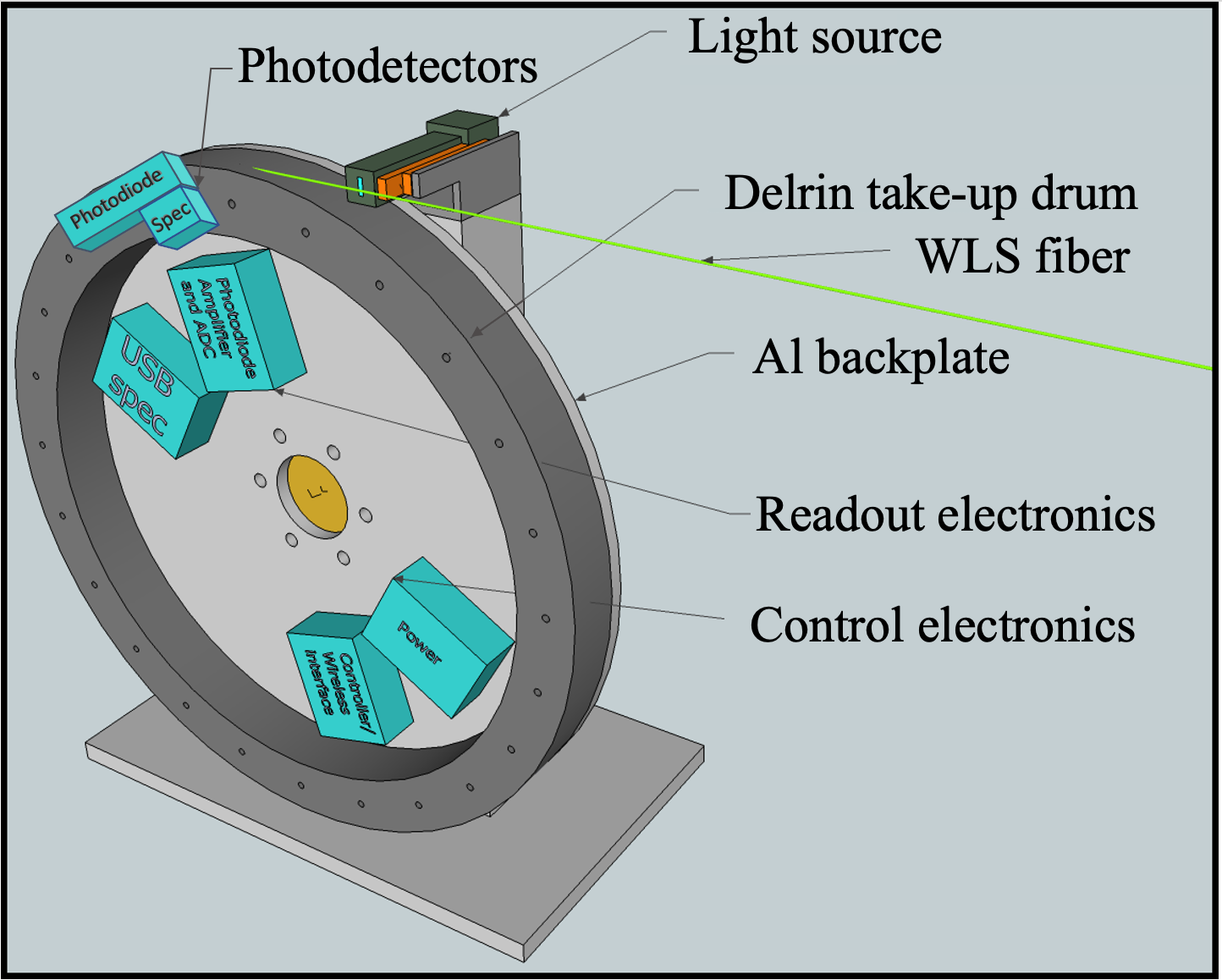}
\includegraphics[height=2.3in]{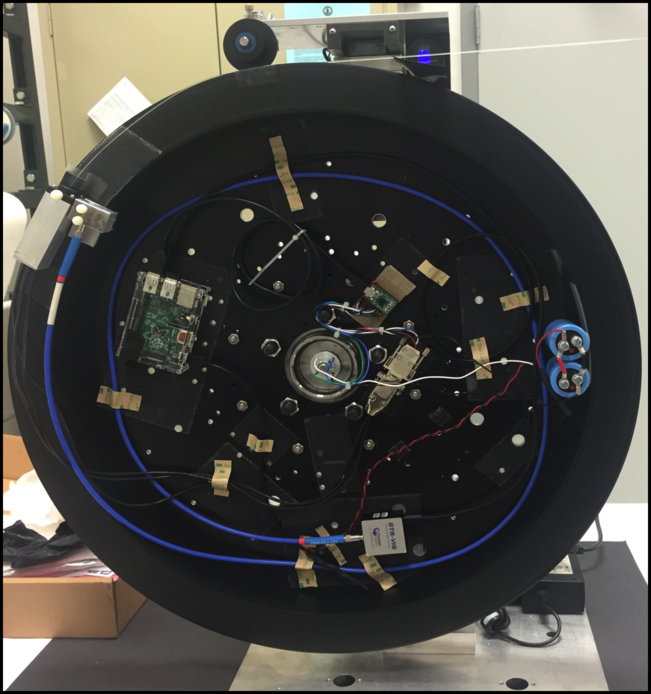}
\caption{A schematic (left) and picture (right) of the fiber scanner.}
\label{fig:scanner}
\end{figure}

A fiber scanner (figure~\ref{fig:scanner}) was designed and fabricated\footnote{Fiber scanner design is by D. Shooltz of Shooltz Solutions LLC.} to study fiber quality for the CRV detector. It measures the WLS fiber light yield as a function of wavelength by varying the distances between a light source and a receiver (either a spectrophotometer or a photodiode). It consists of a 62\,cm diameter take-up drum holding the optical readout electronics, to which the fiber end is connected. The first 25\,m of fiber is wound onto the drum's grooved outer rim. A blue light-emitting diode (LED) source~\cite{led} not affixed to the rotating take-up drum excites the fiber, and the resulting light is captured and readout by a large-area Hamamatsu S1227-1010BR photodiode~\cite{photodiode} or a STS-VIS USB Ocean Optics spectrophotometer~\cite{sts}, depending on the desired measurement. The take-up drum stops at regular intervals to measure the response from the blue LED source at different distances along the fiber, thus providing light yield measurements as a function of wavelength and distance from the readout end. Sample light yield responses from the photodiode and spectrometer are presented in the following. 

The fiber is delivered and stored on 90~cm diameter cardboard spools containing 0.5~km of fiber. The fiber remains on the spool during the measurement, which takes place in a blue-light filtering dark room to minimize fiber damage from harmful UV light. The fiber end is epoxied into a ferrule, polished with a diamond-bit flycutter, then connected to either the photodiode or spectrometer port affixed on the rotating take-up drum. A stepping motor drives the take-up drum with controlled gentle acceleration and deceleration. The drum is moved in 50 cm increments over the first 25\,m of fiber for each spool, and light spectra and intensity measurements are performed at each point. These measurements are assumed to be sufficient to determine the quality of the entire spool. The fiber scan procedure is controlled remotely via WiFi through a web interface on a Raspberry Pi~\cite{pi} mounted on the take-up drum. During data collection, the Raspberry Pi controls the stepper motor and gathers and stores the photodiode or spectrometer data. The scanner is described in more detail in reference~\cite{Dukes:2018scs}.

\section{Results}

\subsection{1.8~mm production fiber}

The CRV detector consists of 5344 counters and 52.3\,km of WLS fiber. The original 1.4\,mm diameter fiber was delivered on 104 spools in May 2018 and all spools were tested with the fiber scanner by the end of June 2018. For improved background rejection, the fiber is being upgraded to larger 1.8\,mm diameter fiber in the most critical regions of the CRV.  The fabrication of the modules in these regions was scheduled at the end of production so they could be built with maximum knowledge and fabrication experience. This larger diameter fiber was delivered on 34 spools in May 2021 and all spools were tested with the fiber scanner by mid-June 2021.\footnote{Fiber diameters were also measured for all 34 1.8\,mm diameter fiber spools as part of the quality control procedure.  A mean with standard error of $1.801 \pm 0.002$\,mm was obtained.} 

\begin{figure}[htb]
\centering
\includegraphics[width=2.95in]{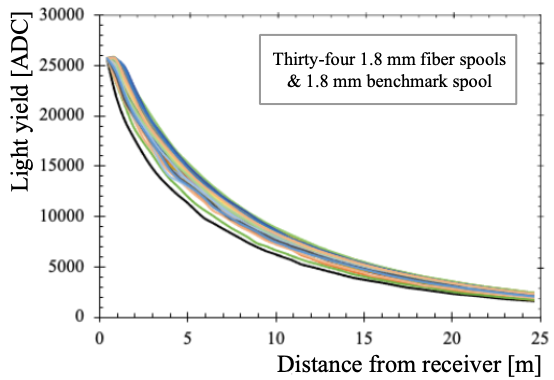}
\includegraphics[width=2.95in]{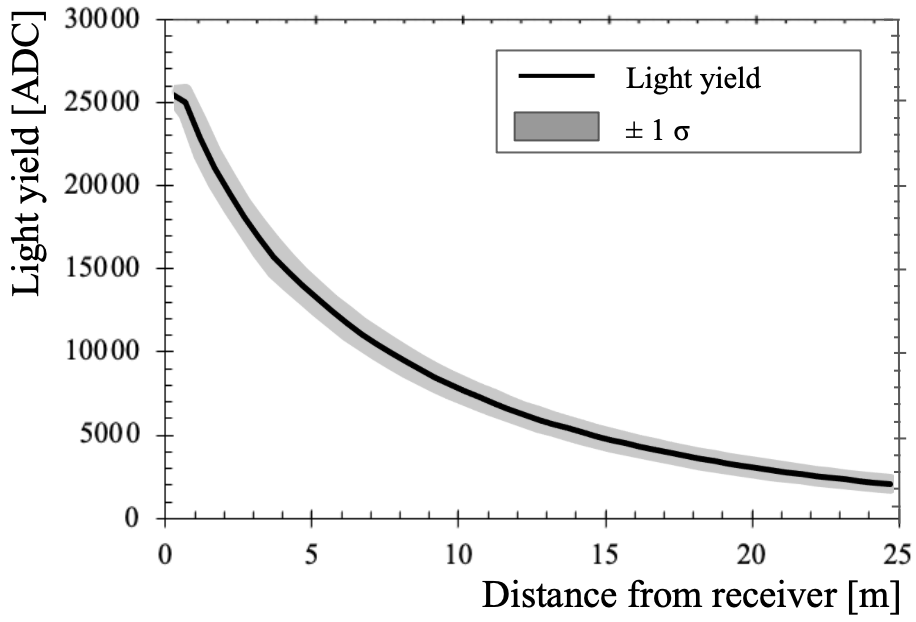} 
\caption{Light yield as a function of the length of the fiber between the source and readout for all 1.8~mm diameter spools (left) and their average (right). The lowest black line on the left is the 1.8\,mm benchmark spool. Measurements are taken in 50 cm increments over the first 25 m of each fiber spool. All spools achieved the desired light yield performance and showed a significant improvement over the original 1.4 mm spools. All spools demonstrate the typical rapid fall in light yield at short distances followed by a less rapid decline.}

\label{fig:ProductionYield}
\end{figure}

\begin{figure}[htb]
\centering
\includegraphics[height=2.3in]{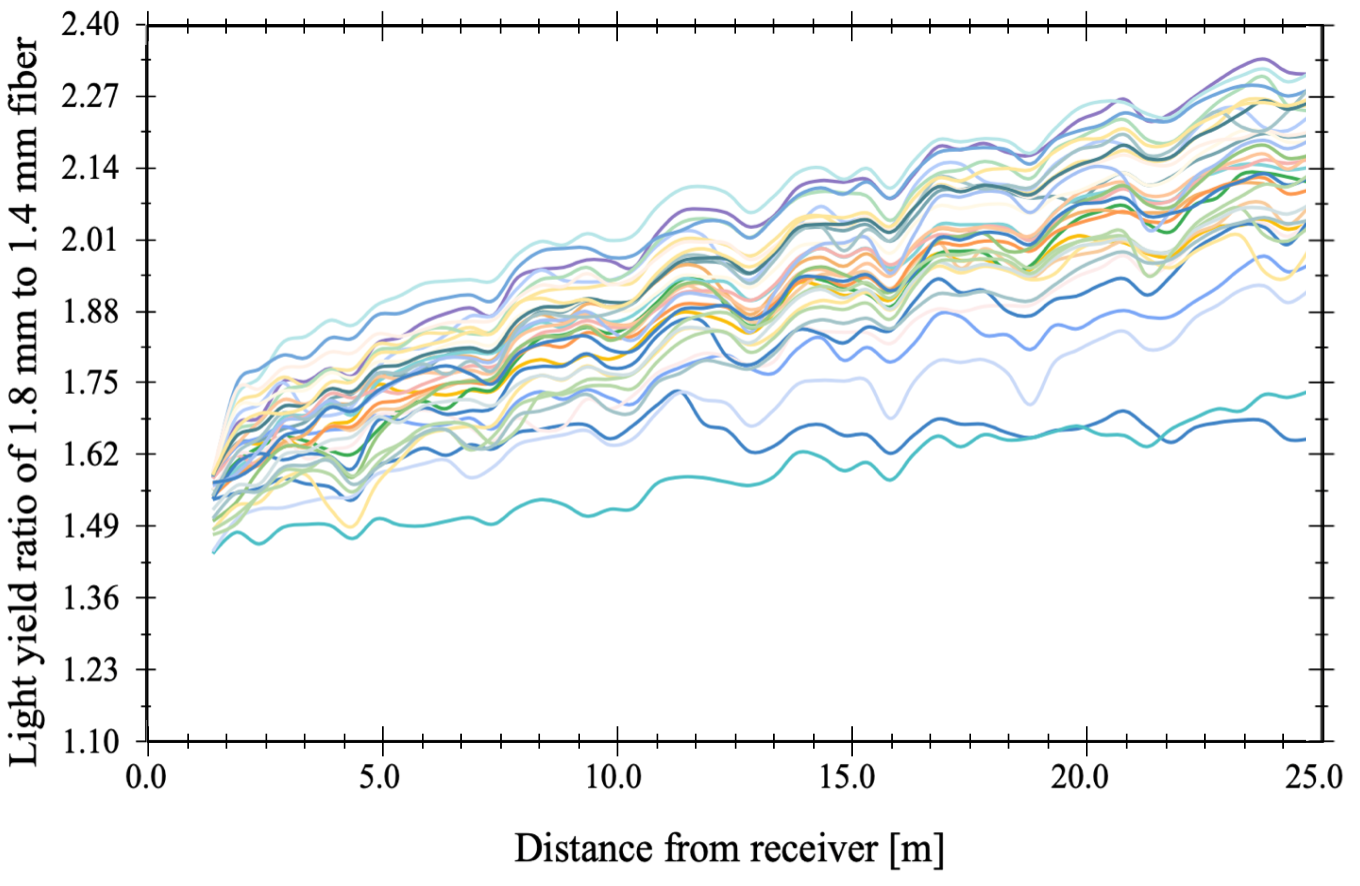}
\caption {Light yield as a function of the distance from the LED source for all thirty-four 1.8\,mm spools shown as a ratio to an original 1.4\,mm fiber manufactured in 2017 and re-tested in June 2021. All tested fibers exceeded the 1.4\,mm fiber's light yield performance. The measurements of the two spools with the lowest response were confirmed with repeated measurements.
}
\label{fig:ProductionYieldRatio}
\end{figure}

The 1.8\,mm production spools light yield is shown as a function of the length of the fiber between the photo-cathode and the LED source in figure~\ref{fig:ProductionYield}, with each curve representing a different fiber spool. This figure also includes a 1.8\,mm diameter fiber spool from 2018 that was re-measured in 2021 (referred to as the benchmark spool). In figure~\ref{fig:ProductionYieldRatio} each spools light yield is normalized to a 1.4\,mm diameter fiber spool from the initial 2018 fiber batch, which was then re-tested in 2021. On average, the production spools light yield response shows a 60\%-70\% increase over the original 1.4\,mm fiber and a 30\% increase with respect to the benchmark spool. This unexpected increase in light yield is not observed in measurements of fully constructed di-counters (as reported section \ref{sec:compare}). The light yield increases can be explained by a combination of aging effects (discussed in section \ref{sec:aging}) and an increase in fiber diameter.

\begin{figure}[htb]
\centering
\includegraphics[width=2.95in]{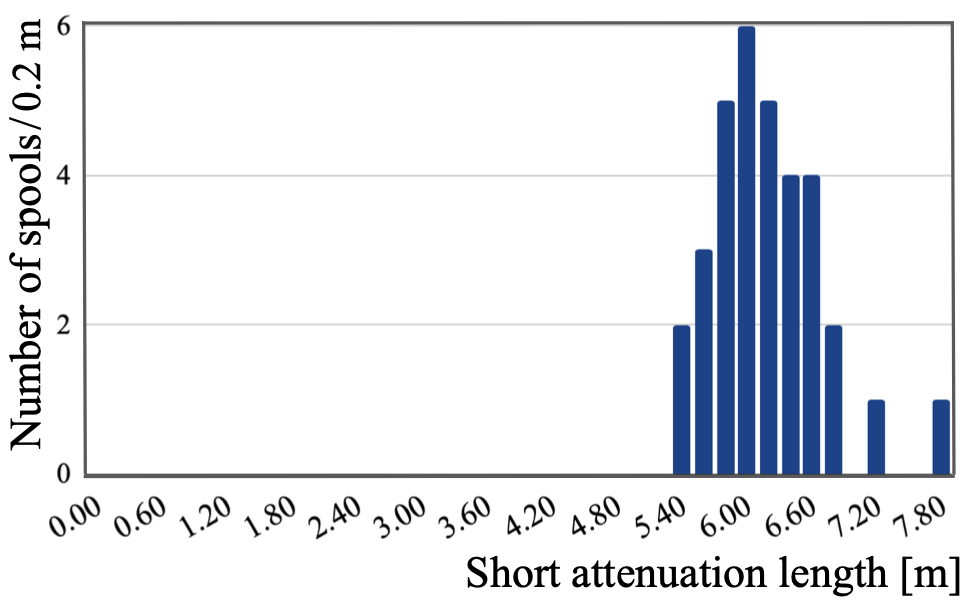}
\includegraphics[width=2.95in]{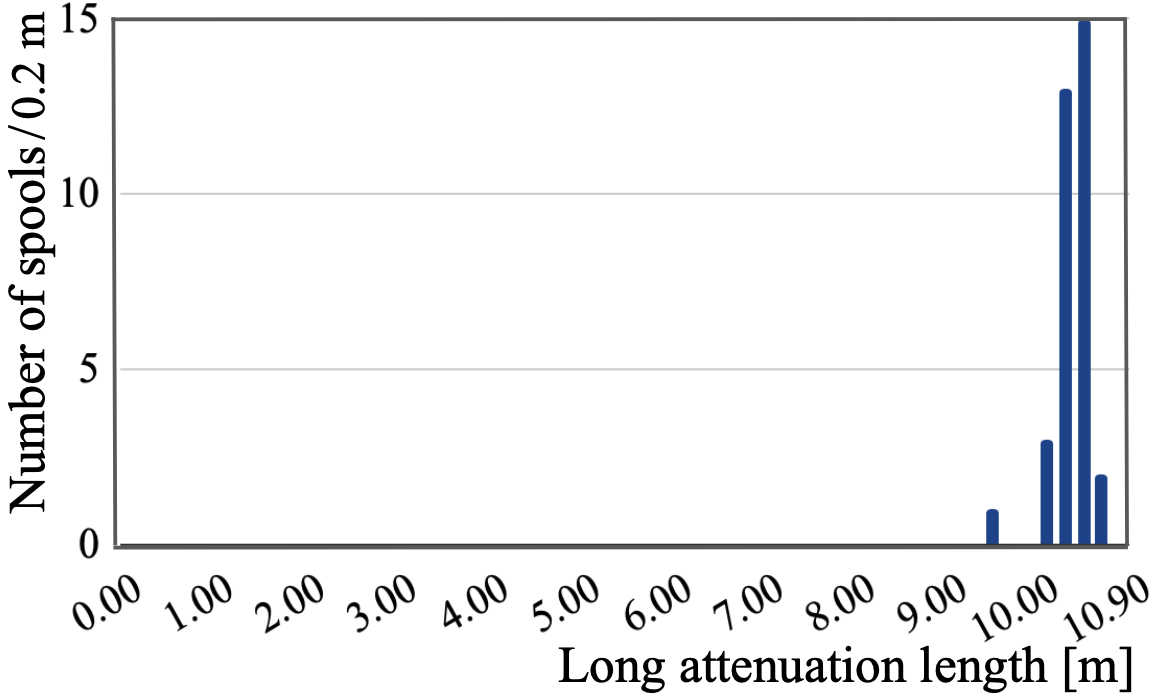}
\caption{The distribution of short (left) and long (right) attenuation lengths for the thirty-four 1.8\,mm spools with a bin width of 0.2\,m. The short attenuation length mean and standard deviation are $6.3 \pm 0.5$\,m, while the long attenuation length statistics are $10.4 \pm 0.2$\,m.}

\label{fig:Ratio}
\end{figure}

\newpage
Short and long attenuation lengths are quantified by separating the data into two distinct categories by fitting an exponential to the range 1-3\,m and 3-25\,m from the fiber end, respectively, and extracting the exponential constants. Distributions of fiber short and long attenuation lengths, as measured from the readout end, are shown in figure~\ref{fig:Ratio}. 

\begin{figure}[htb]
\centering
\includegraphics[width=3.0in]{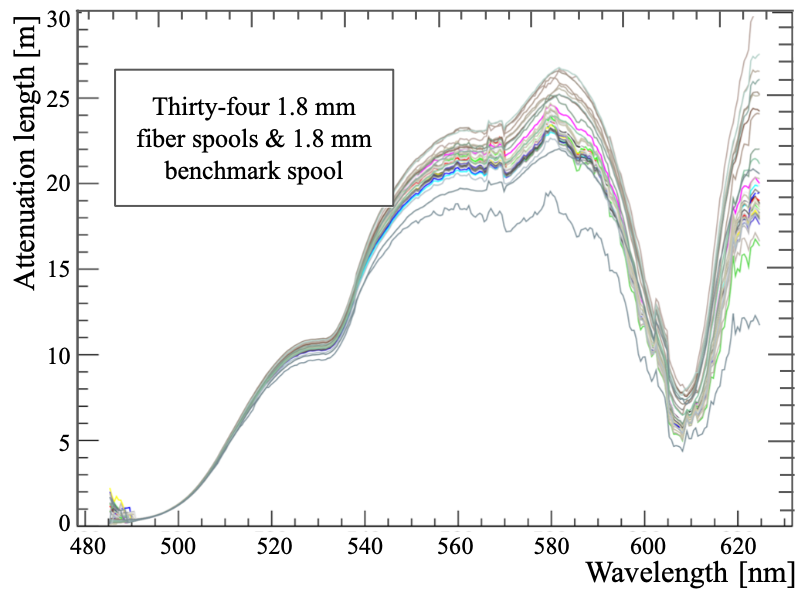}
\caption{Attenuation length as a function of fiber emission wavelength for all production spools and the benchmark spool (lowest curve). These attenuation lengths are extracted using a single exponential fit to the spectrometer data taken from the first 25\,m of each fiber spool. 
}
\label{fig:ProductionSpectrum}
\end{figure}

The spectrometer enables light yield measurements, separated by wavelength in bins of 1 nm, at 50\,cm increments along the first 25\,m of fiber. A single attenuation length is extracted for each wavelength using an exponential fit of the spectral light intensity as a function of distance from the readout end. This single exponential function yields a reasonable fit over the measured 25\,m of fiber when applied to single spectral measurements. Figure~\ref{fig:ProductionSpectrum} shows the attenuation lengths for all production fiber and the benchmark spool as a function of wavelength. The attenuation length drop around 610\,nm is due to an absorption resonance from the fiber's polystyrene core while the attenuation lengths around the optimal SiPM sensitivity range show low variance. Overall, all new 1.8~mm production fibers displayed higher light yield responses and longer attenuation lengths than the benchmark spool across the wavelength ranges. 

\subsection{S-type versus non-S-type fiber comparison} 

\begin{figure}[htb]
\centering
\includegraphics[width=2.85in]{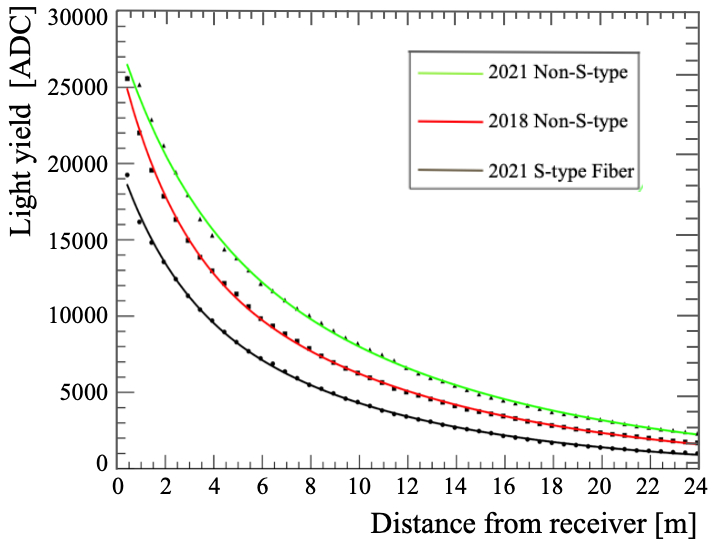}
\includegraphics[width=3.0in]{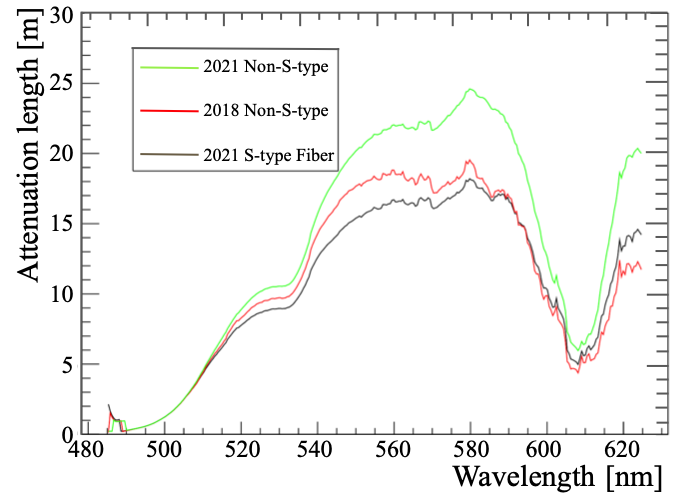}
\caption{Comparison of 1.8\,mm S-type and non-S-type fiber light yield performance (left) and attenuation length vs wavelength (right) for three fibers manufactured in the displayed years. The Mu2e experiment uses non-S-type fiber while the 1.8\,mm LDMX fiber was S-type fiber.}
\label{fig:ComparisonSpectrum}
\end{figure}

There are two fiber types considered in this work, S-type and non-S-type, whose differences arise from the presence or absence of a molecular core (Polystyrene chain) orientation, respectively. Mu2e uses non-S-type fibers due to its greater transparency (more than 10\% longer attenuation length in comparison to S-type fiber), but a sample 1.8\,mm S-type fiber was purchased and tested for a hadronic calorimeter (HCal) prototype for the Light Dark Matter eXperiment (LDMX)~\cite{LDMX}. A minor inconvenience of the non-S-type fiber is a lower flexibility, which requires additional precaution during fiber handling. Figure~\ref{fig:ComparisonSpectrum} shows a comparison of the S-type fiber purchased for the LDMX HCal prototype with the 1.8\,mm non-S-type spools purchased for the Mu2e CRV. As expected, greater light yield and attenuation-length performance is observed for the non-S-type fiber.


\subsection{Fiber light yield comparison}\label{sec:compare}

As mentioned earlier in the paper, light yield measurements were conducted in the Fermilab test beam to determine the light yield gain from larger fiber diameter for fibers of the same age. A 24\% light yield gain was measured for the 1.8\,mm fiber over the original 1.4\,mm fiber. During the fiber testing runs at the University of Virginia (UVA), a sample di-counter was created by threading 1.4\,mm diameter fiber through one scintillator side and 1.8\,mm through the other side and tested with UVA's cosmic test stand \cite{boithesis} to determine the actual light yield increase expected in the CRV sectors. The larger diameter fiber once again recorded about a 25\% light yield increase over the 1.4\,mm fiber which is consistent with geometrical expectations. In addition, several fully constructed CRV modules with the new 1.8 mm fiber also show a similar light yield increase. Both the test-beam data and measurements from the cosmic ray test stand confirm that larger fiber diameter achieves the expected gain in light yield. However, the larger than expected light yield increase as measured by the fiber scanner, beyond what was measured in the fully constructed di-counters, hinted at potential aging effects in the fiber for wavelengths outside of the spectral sensitivity range of the SiPMs. This is described in the following section.

\subsection{Fiber aging studies}\label{sec:aging}

After observing that the benchmark 1.8 mm fiber spool, fabricated in 2015, had a significant decrease ($\sim 30\%$) in light yield in comparison to the new spools from 2021, and that all of the production 1.8 mm spools had a larger than expected increase in light yield over the older 1.4 mm fiber ($\sim 60\%-70\%$), it was decided to investigate potential fiber aging effects. Specifically, five of the thirty-four new 1.8 mm fiber spools, and the benchmark 1.8 mm fiber spool, were remeasured after approximately one year with both the spectrometer and photodiode to provide a direct measurement of aging effects.\footnote{Only five spools were available for this measurement since the other spools had already been used for CRV production.} 

\begin{figure}[htb]
\centering
\includegraphics[width=4in]{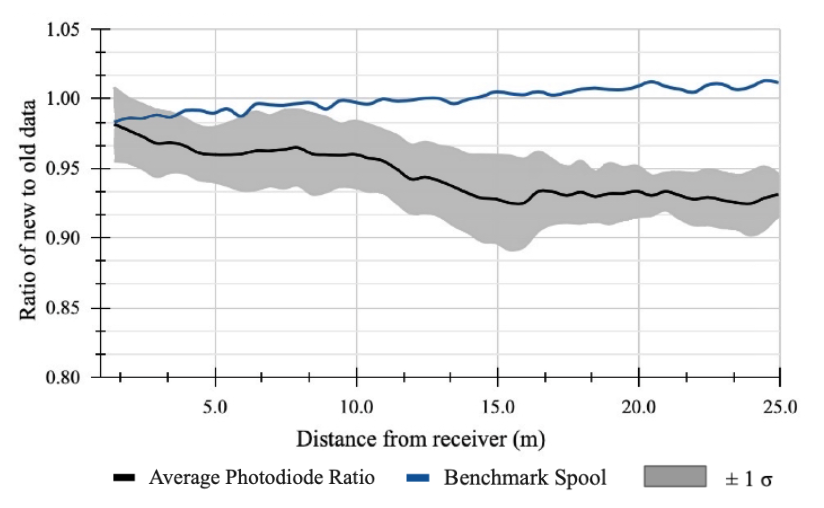}
\caption{The average ratio of new to old photodiode measurements taken approximately one year apart for the five new 1.8\,mm spools (black) and the benchmark spool (blue). There is some apparent aging for the new spools, while there is no measurable aging in the benchmark spool. This indicates that there may be aging effects occurring in the first few years of the fiber's lifetime. The shaded region represents the standard deviation of the five measured spools.}
\label{fig:aging}
\end{figure}

The results, shown in figure~\ref{fig:aging}, indicate a few percent decrease in light yield in the newer spools but no appreciable aging effects in the benchmark spool. This hints that aging might occur within the first years of a fiber's lifetime, but slows down significantly in later years. In addition, the increase in light yield loss with the distance from the readout end points to a decrease in attenuation length. From figure~\ref{fig:ProductionSpectrum}, there are hints of decreasing attenuation lengths for these wavelengths above 520 nm; however, there is also large run-to-run variance for these larger wavelengths which can range between 1-5\%. Thus, narrowing this measured aging effect to a specific wavelength is inconclusive in this work. 

\begin{figure}[htb]
\centering
\includegraphics[width=3in]{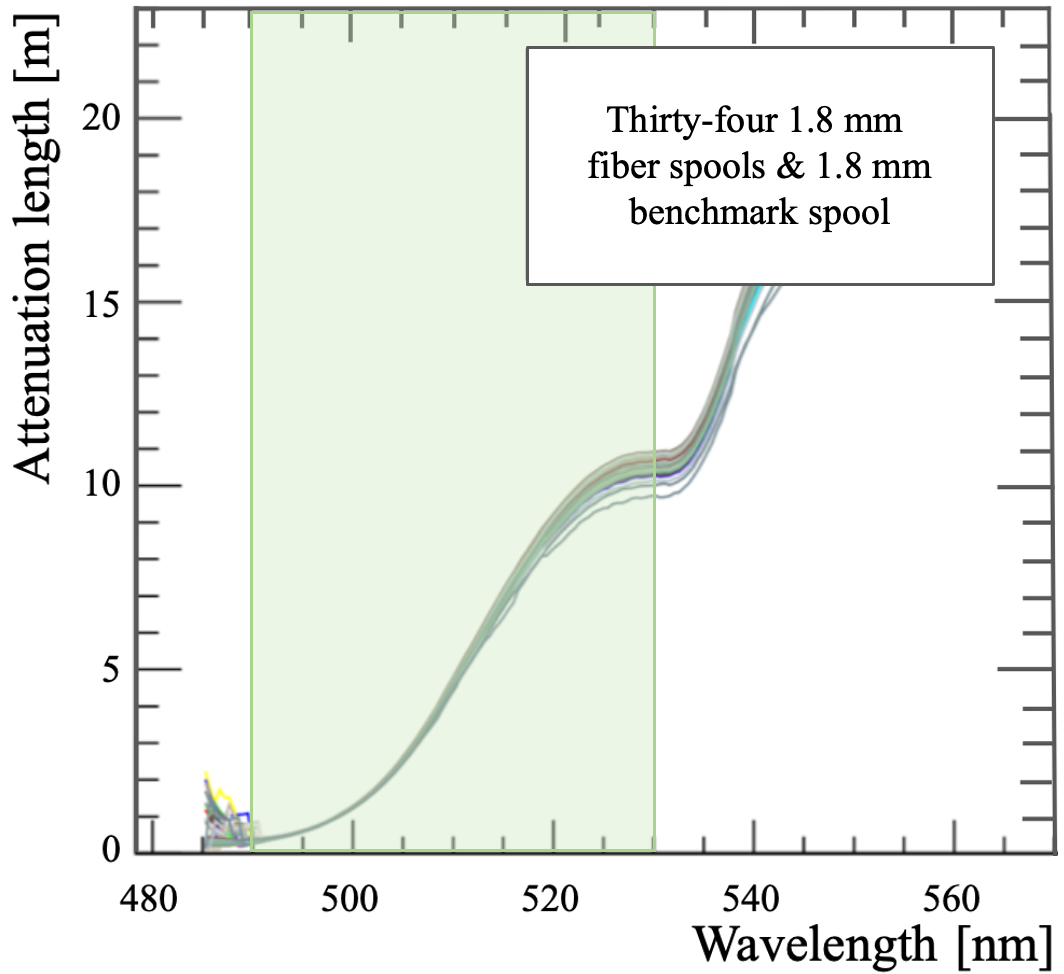}
\caption{A zoomed in version of figure \ref{fig:ProductionSpectrum} in which the shaded region illustrates the wavelengths of interest for the Mu2e CRV. Wavelengths below 490 nm have attenuation lengths much shorter than the length of a Mu2e CRV di-counter. Above 530 nm, the SiPM quantum efficiency and fiber emission spectrum is about 20\% of its peak and drops rapidly.}
\label{fig:aging2}
\end{figure}

Caution should be exercised in interpreting these results in the context of potential aging that may occur within fully constructed Mu2e CRV di-counters. To start, this is not a measurement of the decrease in light yield due to fiber aging within a fully constructed counter as measured by a SiPM. Measurements of light yield for several fully constructed CRV modules do not support rapid fiber aging effects. The difference is that both the spectrometer and photodiode measure a broad spectrum of wavelengths, while the SiPM measurements occur in a much narrower region of wavelength as illustrated in Fig. \ref{fig:aging2}. The measured aging of a fully constructed module in the CRV is still likely dominated by scintillator aging effects as discussed in the previous section. In addition, the longest CRV module is 7\,m while much of the measured aging occurs beyond that length scale. 

It is also possible that, in addition to real aging effects, manufacturing differences between batches of spools produced several years apart could contribute to the apparent aging between the benchmark spool and the new spools. In order to obtain a precise quantitative measure of aging in these fibers, a more careful dedicated study will need to be performed which is beyond the scope and interest of the Mu2e CRV. Specifically, testing many spools from the same manufactured batch multiple times at regular time intervals (on a month to year time scale) using a dedicated setup in a temperature-controlled environment would narrow the possibilities discussed above.

\section{Conclusion}

Light yield and attenuation length measurements were conducted for thirty-four non-S-type and one S-type of 1.8\,mm diameter Kuraray Y11 WLS fiber using a fiber scanner designed for the Mu2e CRV detector. The fibers were purchased to enhance the light yield in the CRV's critical sectors. A significant light yield increase of about 25\% was observed for the 1.8\,mm over the 1.4\,mm diameter fibers. Preliminary measurements report less than 5\% fiber aging per year and suggests negligible aging effects for fibers more than five years old.


\acknowledgments

The bulk of this research was funded by the DOE HEP UVA base grant under DE-FOA-0001781. The S-type fiber was funded with grant number 20190875 from The Crafoord Foundation and purchased for a test-beam prototype for the LDMX experiment. 

This document was prepared by members of the Mu2e Collaboration using the resources of the Fermi National Accelerator Laboratory (Fermilab), a U.S. Department of Energy, Office of Science, HEP User Facility. Fermilab is managed by Fermi Research Alliance, LLC (FRA), acting under Contract No. DE-AC02-07CH11359. Finally, the authors would like to thank Kuraray and their staff for providing the QA measurements and allowing us to discuss them in this paper.  




\end{document}